\documentclass[12pt,3p]{elsarticle}
\usepackage{amssymb}
\usepackage{amsmath}
\usepackage{amsfonts,amsthm,bm} 
\usepackage{graphicx}
\usepackage{caption}
\usepackage{hyperref}
\usepackage{color, colortbl}
\definecolor{silver}{rgb}{0.75,0.75,0.75}
\usepackage{xcolor}
\usepackage{lscape}
\usepackage{rotating}
\usepackage{diagbox}
\usepackage{subcaption}
\usepackage{csquotes}
\usepackage{floatpag}
\usepackage{setspace}
\usepackage{float}
\usepackage{dcolumn}
\usepackage{bbm, dsfont}
\usepackage{ulem}
\PassOptionsToPackage{hyphens}{url}\usepackage{hyperref}

\makeatletter
\def\ps@pprintTitle{%
\let\@oddhead\@empty
\let\@evenhead\@empty
\def\@oddfoot{\centerline{\thepage}}%
\let\@evenfoot\@oddfoot}
\makeatother

\usepackage{etoolbox}
\patchcmd{\MaketitleBox}{\footnotesize\itshape\elsaddress\par\vskip36pt}{\footnotesize\itshape\elsaddress\par\parbox[b][36pt]{\linewidth}{\vfill\hfill\textnormal{\today}\hfill\null\vfill}}{}{}%
\patchcmd{\pprintMaketitle}{\footnotesize\itshape\elsaddress\par\vskip36pt}{\footnotesize\itshape\elsaddress\par\parbox[b][36pt]{\linewidth}{\vfill\hfill\textnormal{\today}\hfill\null\vfill}}{}{}

\newcommand{\bc}[1]{{\color{black}{#1}}} 
\newcommand{\CE}[1]{{{}}}
\newcolumntype{L}{D{.}{.}{2,5}}

\begin{document}
\begin{frontmatter}

\renewcommand*{\thefootnote}{\fnsymbol{footnote}}

 \title{ An Information Filtering approach to stress testing: an application to FTSE markets }
\author{Isobel Seabrook, Fabio Caccioli, Tomaso Aste}
\address{Department of Computer Science, University College London, \\ Gower Street,  WC1E 6EA, London, United Kingdom}

\begin{abstract}
We present a novel methodology to quantify the `impact' of and `response' to market shocks.  
We apply shocks to a group of stocks in a part of the market, and we quantify the effects in terms of average losses on another part of the market using a sparse probabilistic elliptical model for the multivariate return distribution of the whole market. 
Sparsity is introduced with an $L_0$-norm regularization, which forces to zero some elements of the inverse covariance according to a dependency structure inferred from an information filtering network.
Our study concerns the FTSE 100 and 250 markets and analyzes impact and response to shocks both applied to and received from individual stocks and group of stocks. 
We observe that the shock pattern is related to the structure of the  network associated with the sparse structure of the inverse covariance of stock returns. 
Central sectors appear more likely to be affected by shocks, and stocks with a large level of underlying diversification have a larger impact on the rest of the market when experiencing shocks. 
By analyzing the system during times of crisis and comparative market calmness, we observe changes in the shock patterns with a convergent behavior in times of crisis.
\end{abstract}


\end{frontmatter}

\section{Introduction}
We present a novel approach to reverse stress testing of markets, which allows identification of stocks that either have the largest impact on other stocks, or show the largest response to shocks from others. 
 We measure stress propagation in terms of the conditional distribution of losses, which we compute by modeling the return distributions of the entire market in terms of a multivariate elliptical probability distribution\cite{aste2020stress}.

Impact is quantified as the average losses caused by a set of stressed stocks on the rest of the system. 
Conversely, response is quantified as the average losses suffered by a set of stocks when the rest of the system is stressed. We apply this method to different states of the FTSE 100 and 200 markets.

The estimation of the multivariate probability is made accurate by using sparse inverse covariance estimation, where the non-zero elements are the edges of an information filtering network.
We observe that the structure of this network is related to the behaviour of the system when stressed. 
In particular, we observe that impact and response of individual nodes are both related to their centrality, but when we consider groups of nodes, the most central groups have a higher response but lower impact due to the large internal effect that the related stocks have on each other. 

With a regression analysis, we investigate if the impact and response measures of industry supersectors are affected by their centrality, and we find that the size and fraction of links shared by the nodes of a supersector are significant for impact, but centrality is not. In contrast, centrality is significant for response, suggesting that more central sectors are more likely to be affected by shocks. 
This is in line with what reported in \cite{pozzi2013spread}, where it was observed that portfolios with stocks belonging to the peripheral region of information filtering networks are less risky.  However, here we quantify this risk in terms of propagation of stress across the market, providing a tool to hedge risk and identify vulnerabilities within a reverse stress-testing framework.

Finally, we show how the information filtering network can be used to reverse stress test the system. More specifically, we identify the group of $10$ nodes that collectively have the largest impact on the system. We find that these nodes correspond to funds, suggesting that stocks with a high level of underlying diversification have a higher propensity to impact the rest of the market. 

By using the sparse probabilistic model of the whole market, we extract six temporal clusters that represent periods with different market behaviour. The different clusters are identified from the similarity/dissimilarity in the likelihoods of the daily set of returns across all stocks (see ICC algorithm in \cite{procacci2019forecasting}).  
We observe that there are significant differences in the different market states, with periods of crisis showing a convergence of group behavior to the single node trend, indicating,  in line with \cite{Leonidas_2011}, convergent behavior in times of crisis. 

We consider our approach to be a form of ‘reverse stress testing’ in a macroprudential sense. A strict microprudential definition of ‘reverse stress testing’ describes it as an exercise that involves exploring the size and nature of shocks that would render a bank’s business model unviable, or its financial position fragile. It starts from an outcome of business failure and identifies circumstances where this might occur \cite{BOEFSR2020}. For the interested reader we point out that a summary of  state-of-the-art approaches on MaPST, and roadmap for future research, was presented by the IMF in \cite{IMF_stress}.

\renewcommand*{\thefootnote}{\arabic{footnote}}
\setcounter{footnote}{0}

\onehalfspacing

\section{Background}
This paper builds upon several methodologies that have been developed in recent years, concerning four main areas:
\begin{enumerate}
    \item Modeling markets in terms of multivariate probability distributions;
    \item  Use of networks to describe the interrelations between the variables in markets;
    \item Sparsification of the probabilistic models using these networks; 
    \item Identification of different states of the market associated with different sparse multivariate probability distributions. 
\end{enumerate}

Hereafter, we briefly summarize the necessary state-of-the-art background in these area.

\subsection{Modeling markets with multivariate elliptical distributions}
The elliptical distribution family is a broad family of multivariate probability density functions that, notably, includes the multivariate normal and the multivariate Student-t distributions. In this paper\bc{,} we model the multivariate probabilistic structure of the log-returns for the entire market in terms of a multivariate elliptical probability distribution. We shall see shortly that, for the measure have we chosen as quantification of stress impact, there is no need to specify the kind of distribution within this broad family. Within this multivariate probabilistic framework, stress is naturally modelled in terms of the multivariate conditional probability. Specifically, the probability distribution of the stressed variables is conditioned on the values of the stressing variables \cite{aste2020stress}.  The challenge is to estimate accurately the inference structure and the parameter values of such  multivariate probability distribution.

\subsection{Information filtering networks}

Information filtering networks are particularly effective for correlation-based graphs, and it has been shown that they describe well the structure of the market and its evolution with time \cite{tumminello2005tool,aste2010correlation}. 
They have been used to characterise the spread of risk across a market \cite{pozzi2013spread}, and it has been shown that risk does not distribute uniformly and the central or peripheral position of an asset in the market is an important risk factor. 

It is common in the literature to find studies which link network centrality and systemic risk\cite{markose2012too,battiston2012debtrank,martinez2014empirical,bravo2016centrality}, and this has been exploited by policy makers in order to inform systemic risk rankings \cite{bruyckere2015ecb}. A direct focus on centrality measures in detecting systemically important financial institutions was applied by Kuzubaş et. al. \cite{kuzubas2014turkishfc}, confirming ex-post that centrality values perform well in detecting systemically important institutions in interbank markets.

More general aspects of network structure have also been considered, for instance Billo et al. \cite{billio2012econometric} show that indirect measures of firm interconnectedness based on principal component analyses and granger-causality networks are able to indicate periods of market dislocation and distress. In a similar vein, Diebold and Yilmaz \cite{diebold2014topology} propose connectedness measures at all levels from system-wide to pairwise, and link this to Marginal Expected Shortfall and Conditional Value at Risk to emphasise the usefulness of these measures in a risk measurement and management setting. 

\subsection{Sparse modeling with information filtering networks}
Following the approach proposed in \cite{LoGo16} and \cite{aste2020topological}, we construct sparse probabilistic models that have the information filtering networks encoded in the structure of the inverse covariance (i.e. partial correlations) \cite{LoGo16}. 
Such a method is based on a special family of chordal networks which are clique-forests \cite{massara2019learning}. Specifically, we use the Triangulated Maximally Filtered Graph (TMFG) \cite{massara2016network}, which is one instance of such a family. It is a 3-clique structure made of tetrahedra separated by triangles, and it has the advantage to be computationally very efficient. 
It has been proven to be effective in identifying relevant data structures in different contexts from finance to psychology \cite{nicola2020information,turiel2020simplicial,christensen2018network}. 

These networks can be used to estimate the maximal likelihood solution of elliptical multivariate distributions with sparse inverse covariance. This is a special solution for $L_0$-norm regularization, and it is a valid alternative of the popular Glasso method \cite{friedman2008sparse} for the covariance selection problem.  
Sparsification results in a better estimate of the covariance and overcomes issues related to the curse of dimensionality \cite{aste2020topological}.

\subsection{Identification of different market states}
In this paper we use a recently developed time-clustering methodology (Inverse Covariance Clustering, ICC, \cite{procacci2019forecasting}) that combines the information filtering network description with a complete probabilistic modeling of the system to identify temporal clusters well represented by the same multivariate probability distribution. This method was chosen over other methods, as it efficient and reliable in identifying and predicting accurate and interpretable structures in multivariate, non-stationary financial datasets. This is in contrast to time series models such as TAR \cite{tong1978TAR}\bc{,} which are often unable to identify structural breaks, and time series clustering techniques \cite{ren2017pattern, nevill1997hierarchical, liao2005dtw}\bc{,} which are highly susceptible to the curse of dimensionality. In \cite{procacci2020covidstates}, which considers equities traded in the US market, the method is able to distinguish a market state associated with both the 2008 crisis period and the COVID-19 as a distinct state from the long `bull' period post 2008. ICC was also applied in a recent note in relation to the COVID-19 pandemic \cite{procacci2020covidstates} to identify inherent market structures.

In this work, we similarly identify distinct states for the 2008 period, but for equities making up the FTSE 100 and 250. 
Pharasi et. al. \cite{pharasi2020states} identify market states as clusters of similar correlation matrices applied to stocks making up the S\&P 500 and Nikkei 225 indices. Similar correlation\bc{-}based methods were applied by Münnix et. al.\cite{munnix2012states}, who identify points of drastic change in correlation structure, which map to occurrences of financial crises. An alternative approach has been presented by Hendricks et al. \cite{hendricks2020potts}, in which a maximum likelihood approach is applied to a physical analogy of the ferromagnetic Potts model at thermal equilibrium to cluster temporal periods as objects based on market microstructure feature performance.

\section{Methodology} \label{s.2}

\subsection{Conditional probability measure of systemic risk}
From a probabilistic perspective the quantification of stress contagion between two sets of assets is measured by the conditional probability distribution, $P(\mathbf Y | \mathbf X= \mathbf x)$, of the set of the stressed variables, $\mathbf Y$, under the condition that the set of stressing variables $\mathbf X$ is constrained at a given value of stress $\mathbf x$.
Generally speaking, the conditioning can change both the kind of distribution and its parameters. 
For the vast elliptical distribution family, conditioning at $\mathbf X=\mathbf x$ causes a shift in the expected value of the conditioned variable.
Here, we quantify stress contagion in terms of average loss in a set of assets caused by a loss imposed on another set of assets.  

Consider the losses in the two sets of assets as represented by  two multivariate sets of variables $\mathbf{X} \in \mathbb R^{p_{\mathbf X} \times 1}$ and $\mathbf{Y} \in \mathbb R^{p_{\mathbf Y}\times 1}$. 
Assuming they belong to the multivariate elliptical family probability distribution, then the conditional expected values are
\begin{equation}
\mathbb E [ \mathbf Y | \mathbf X=\mathbf x ] =
\boldsymbol{\mu}_{\mathbf Y} +  \mathbf{\Omega}_{\mathbf{YX}} \mathbf{\Omega}^{-1}_{\mathbf{XX}} (\mathbf{x}- \boldsymbol{\mu}_{\mathbf X})
\label{MuY|X}
\end{equation}
Where $\boldsymbol{\mu}_{\mathbf X}$ and $\boldsymbol{\mu}_{\mathbf Y}$ are the vectors of expected value\bc{s} of the variables $\mathbf X$ and $\mathbf Y$ respectively.
The terms $ \mathbf{\Omega}_{\mathbf X \mathbf X} $, $ \mathbf{\Omega}_{\mathbf Y \mathbf X}$ are the block elements of the  shape matrix $ \mathbf{\Omega}$
\begin{equation}
 \mathbf{\Omega} = 
\left(\begin{array}{cc}
 \mathbf{\Omega}_{\mathbf X \mathbf X}& \mathbf{\Omega}_{\mathbf X \mathbf Y}\\
 \mathbf{\Omega}_{\mathbf Y \mathbf X}&  \mathbf{\Omega}_{\mathbf Y \mathbf Y}
\end{array}\right),
\end{equation} 
where $ \mathbf{\Omega}_{\mathbf X \mathbf X} $ is assumed invertible. 

From \ref{MuY|X}, we observe that the effect of conditioning is shifting the centroid of the variables $\mathbf Y$ by $\mathbf{\Omega}_{\mathbf{YX}} \mathbf{\Omega}^{-1}_{\mathbf{XX}} (\mathbf{x}- \boldsymbol{\mu}_{\mathbf X})$.
It has been argued in \cite{aste2020stress} that such a shift is a good measure of systemic risk quantifying the average losses on variables   $\mathbf Y$ when variables $\mathbf{x}$ deviate from the mean. 
We measure the mean loss in variables  $\mathbf Y$ when $\mathbf{x}- \boldsymbol{\mu}_{\mathbf X}=\mathbf{1_X}$ with the quantity:
\begin{equation}
L_{\mathbf{X \to Y}} =  \frac{1}{  p_{\mathbf Y} } \mathbf{1_Y^\top} \mathbf{\Omega}_{\mathbf{YX}} \mathbf{\Omega}^{-1}_{\mathbf{XX}}\mathbf{1_X}.
\label{e.L}
\end{equation} 
We shall refer to this quantity as `impact'\bc{,} meaning that it qualifies the effect on the rest of the system ($\mathbf Y$) of a unitary stress applied on the group of variables $\mathbf X$.
Similarly, we shall refer to $L_{\mathbf{Y \to X}}$ as `response'\bc{,} and it quantifies the effect on $\mathbf X$ of a unitary stress applied on the rest of the system ($\mathbf Y$).

 \subsection{Identification of most impactful elements}
 Using the above measure. we investigated the group of most impactful variables by devising a simple algorithm that starts from a random group of $n$ variables and then iteratively tries to replace one variable in the group with an external variable not in the group if such a replacement  increases the total impact of the group. The procedure ends when it converges to a stable cluster.
This simple procedure is not deterministic, and there are instances when the final result might change. We however verified that, in practice, in almost all cases the same group of variables is selected. 
This simple procedure is not optimized for numerical efficiency, but for the purpose of this paper, where only a few hundreds of variables are involved, the algorithm converges in fraction of seconds on a standard laptop. 

\subsection{Construction of sparse inverse scale matrix using LoGo}
We use the TMFG information filtering network \cite{TMFG} to estimate the $L_0$-norm regularized sparse inverse shape matrix $\mathbf{\Omega}$. 
Within the elliptical family of probability modeling, the shape matrix is proportional to the covariance (here assumed to be defined), and this sparse  inverse is consequently proportional to the partial correlation. 
The zero entries in the sparse  inverse shape matrix are therefore associated with zero conditional correlations. 
Note that however, beside normal models, such a zero conditional correlation does not imply conditionally independent variables. 
The $L_0$-norm regularized maximum likelihood shape matrix for the whole system is estimated from the local estimates of the shape matrices associated with the tetrahedral cliques in the TMFG, following the approach, called LoGo,  outlined in \cite{LoGo16}.
This provides accurate and robust estimates that overcome the issue of the curse of commonality. 
Furthermore, it has been observed in \cite{LoGo16,procacci2021portfolio}\bc{\sout{,}} that the off-sample likelihoods of the LoGo estimate are larger than the one obtained with the full-matrix max-likelihood estimation or other sparsification methods such as Glasso \cite{friedman2008sparse}.   

\subsection{Identification and clustering of market states}
\label{MarketStateCluster}
We identify market states by clustering together days which are described by  similar multivariate probability structures. 
The probabilistic description uses the multivariate elliptical distribution with a sparse inverse scale matrix. 
The clustering procedure follows \cite{procacci2019forecasting}\bc{,} and is performed by starting from six samples composed of randomly gathered days. 
For each of these clusters, $c=1...,6$, we estimate the vector of means $\bm{\mu}_c$ and the sparse inverse covariance $\bm{J}_c=\bm{\Omega}_c^{-1}$. 
Then the following penalized log-likelihood is computed for each day:
\begin{equation}
	\ell_{c,t} = \log \left| \bm{J}_c \right| - (\bm{X}_t- \bm{\mu}_c)\bm{J}_c(\bm{X}_t- \bm{\mu}_c)^T - \gamma \boldsymbol{1} ( \mathcal C_{t-1}\not= c )  \;\;,
  \label{likelihood_firstDef}
\end{equation}
where  $\bm{X}_t=(x_{t,1},x_{t,2},...,x_{t,n})$  is the $n$-dimensional multivariate log-return vector for each day $t$; $\gamma$ is a parameter penalizing state switching and; $\boldsymbol{1} ( \mathcal C_{t-1}\not= c )$, is a penalizer function for cluster discontinuity that returns 1 if the cluster assignment of the observation at time $t-1$, $\mathcal C_{t-1}$, is different from the cluster assignment at time $t$.
The clustering process is updated iteratively re-assigning days to clusters in a way to maximize $\sum_{t,c} \ell_{c,t}$. 
The procedure is made computationally efficient by using a Viterbi path.
The process stops when a maximum is reached.
This process is not deterministic\bc{,} and it normally ends in a local maximum. However, the clustering structure in the different local maxima is usually rather similar. 
We re-run the whole procedure described in this paper for ten times and we observed that, across all ten runs, the final results and conclusions are consistent with those reported hereafter.

\section{Data}
We consider End of Day prices for stocks making up the FTSE 100 and 250 indices, from January 2005 to August 2020. Of these stocks, 231 of the 350 were available across the whole period analysed; those that weren't were not considered. This time period contained two periods of significant market stress - the 2008 financial crisis, and the initial market shocks experienced in response to the COVID-19 pandemic. 
Making use of the Global Industry Classification Standard, these stocks were classified into 11 Sectors. The number of stocks per sector can be seen in table \ref{tab:Sector_counts}. Some stocks in the raw data were labeled with N/A as the sector; upon inspection it was found that these were all some form of financial fund, and so these stocks were relabelled as `Funds'.

\begin{table}
    \centering
    \begin{tabular}{|c|c|}
    \hline
    Sector & Number of stocks\\
    \hline
       Industrials            &   47 \\
Funds             &          43 \\
Consumer Discretionary  &  29 \\
Financials             &   25 \\
Real Estate        &       20 \\
Materials           &      15 \\
Consumer Staples    &      14 \\
Communication Services &   10 \\
Information Technology  &  10 \\
Utilities            &      7 \\
Health Care          &      5 \\
Energy            &         4 \\
\hline
    \end{tabular}
    \caption{Number of stocks per GICS SuperSector}
    \label{tab:Sector_counts}
\end{table}
\section{Results}

\subsection{Market states} 
\begin{figure}[ht]
\begin{center}
\includegraphics[width=0.8\textwidth]{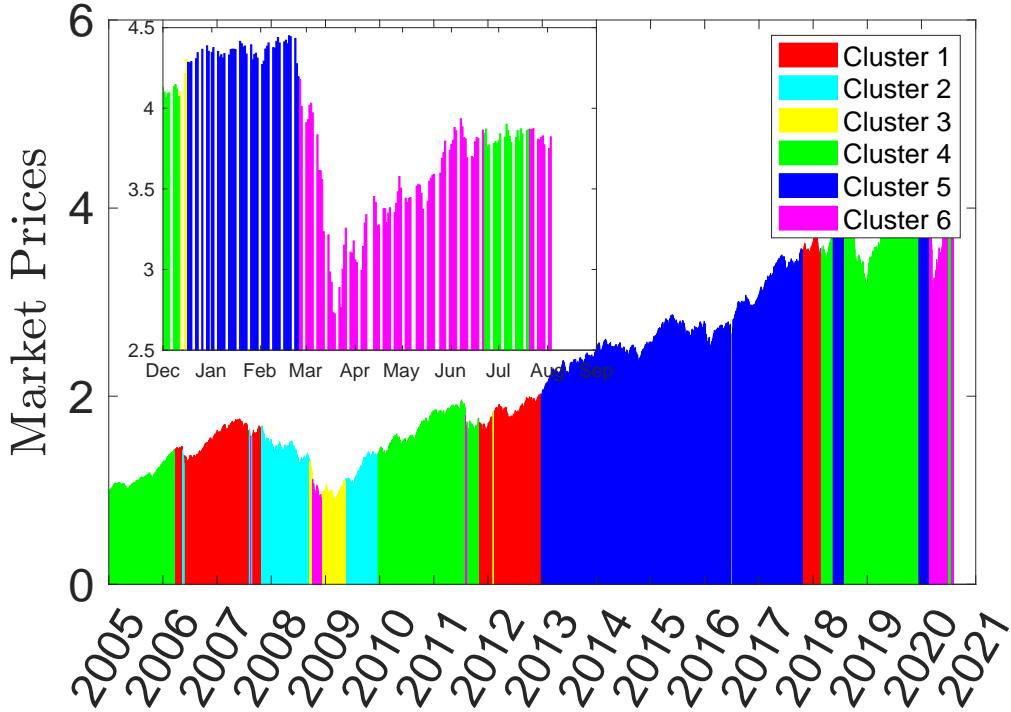} 
\end{center}
\caption{Market states obtained through the clustering algorithm described in Section \ref{MarketStateCluster}.
The y-axis reports the average price across the market (daily data) and the colors represent the six clusters.
}\label{MarketStates}
\end{figure}

The clustering of the system into six market states associated with daily maximal likelihood yields the partition represented in Fig.\ref{MarketStates}.
The choice of six has been mostly guided by the observation that it provides a good distinction between the various periods.
The results have been obtained using $\gamma =100$\bc{,} which provides average sizes of uninterrupted clusters of    about one month. 
Using different values of gamma does not change significantly the results.  
One can observe that the six clusters spread unevenly through the observation window.
The ordering follows their average position in time, with cluster 1 mostly present at the early stages between 2005 and 2013 while cluster 6 is mostly present during the  2020 Covid-related market turmoil. 
Note that we observe days in the 2007-2009 crisis period being also associated with cluster 6\bc{,} however most of that crisis period is associated with cluster 3. 
As already mentioned, this clustering method is not deterministic and different runs can return different clustering. 
Therefore, for this study, the clustering was executed ten times and all results presented in this paper were computed for each instance. Results do not vary significantly and the conclusions are the same for all runs. 

\begin{figure}[ht]
\begin{center}
\includegraphics[width=0.5\textwidth]{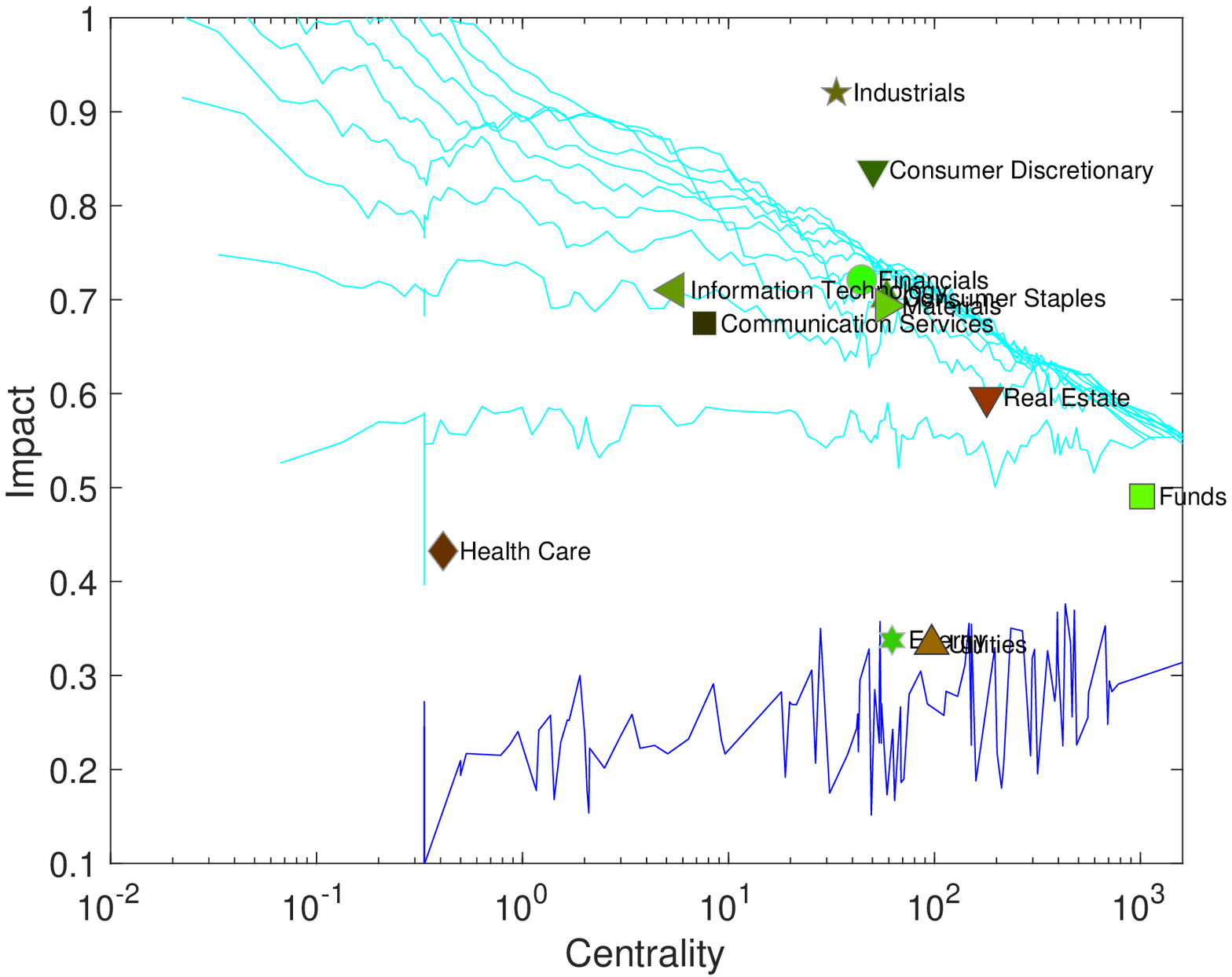}
\includegraphics[width=0.5\textwidth]{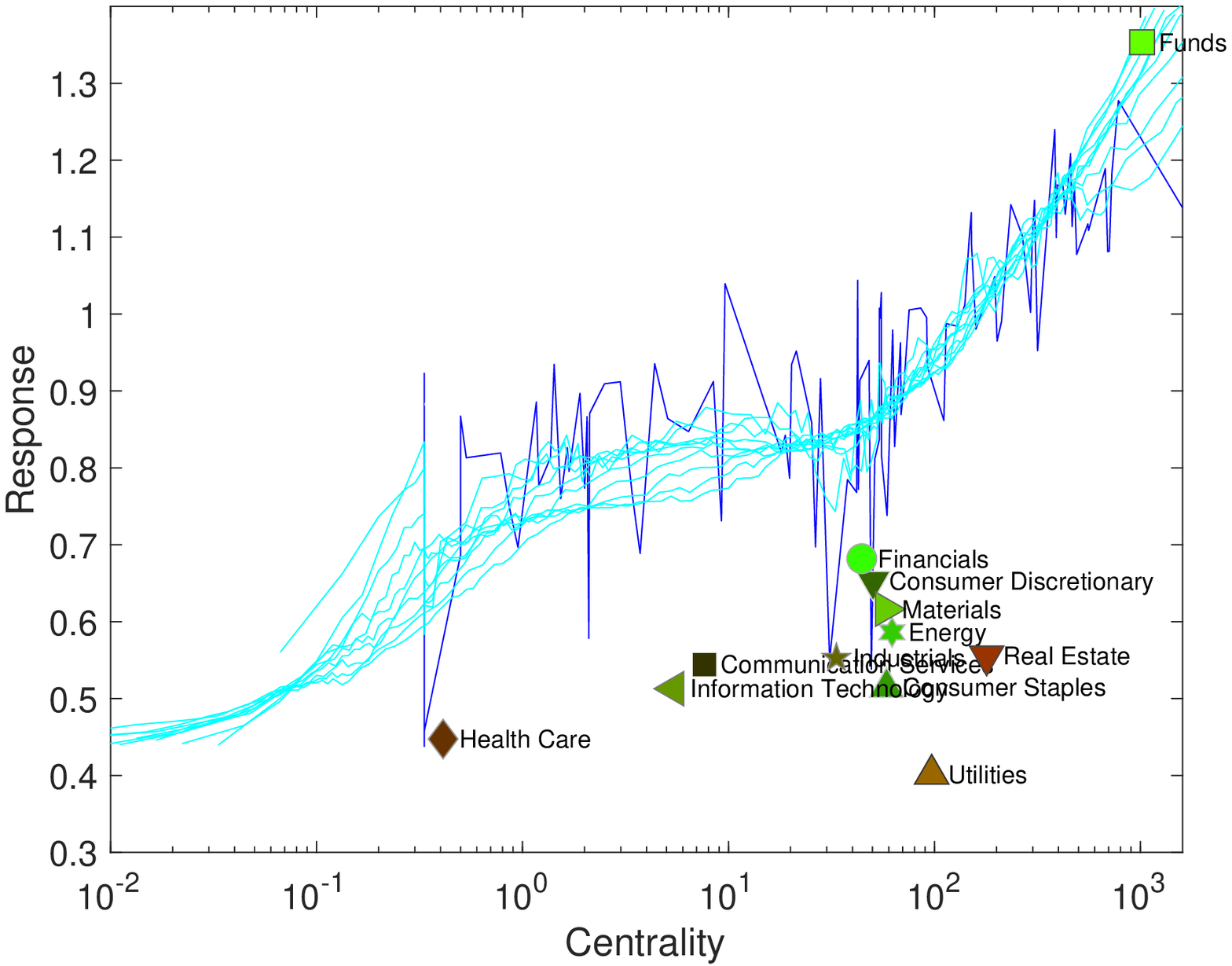} 
\end{center}
\caption{Impact / response vs. centrality computed over the whole period.
The darker blue line correspond to single nodes, cyan lines are group of nodes form 5 to 50.
Symbols represent the supersectors.
}\label{ImpactVSCentraliatyWP}
\end{figure}

\begin{figure}[ht]
\begin{center}
\includegraphics[height=4.cm]{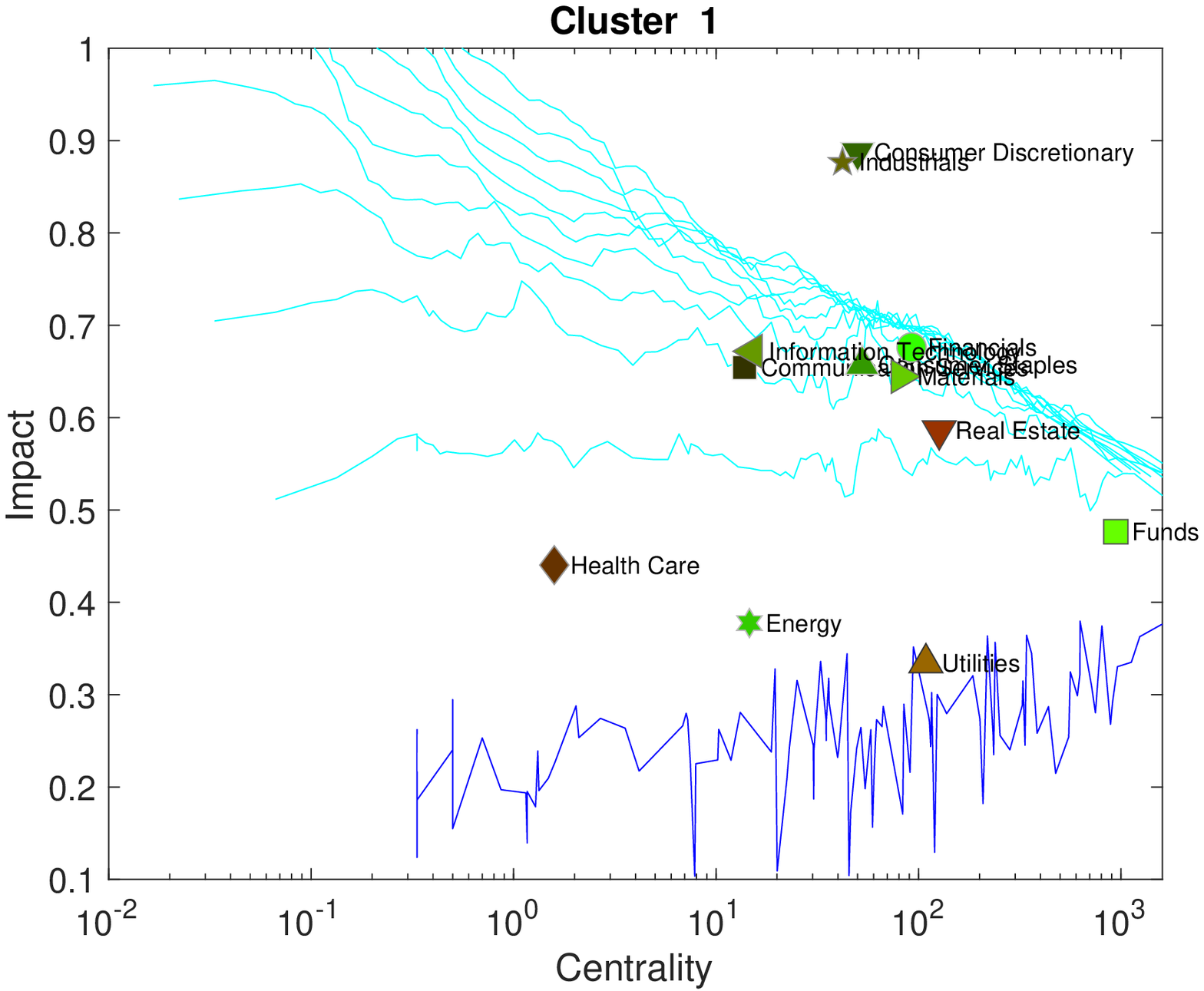}
\includegraphics[height=4.cm]{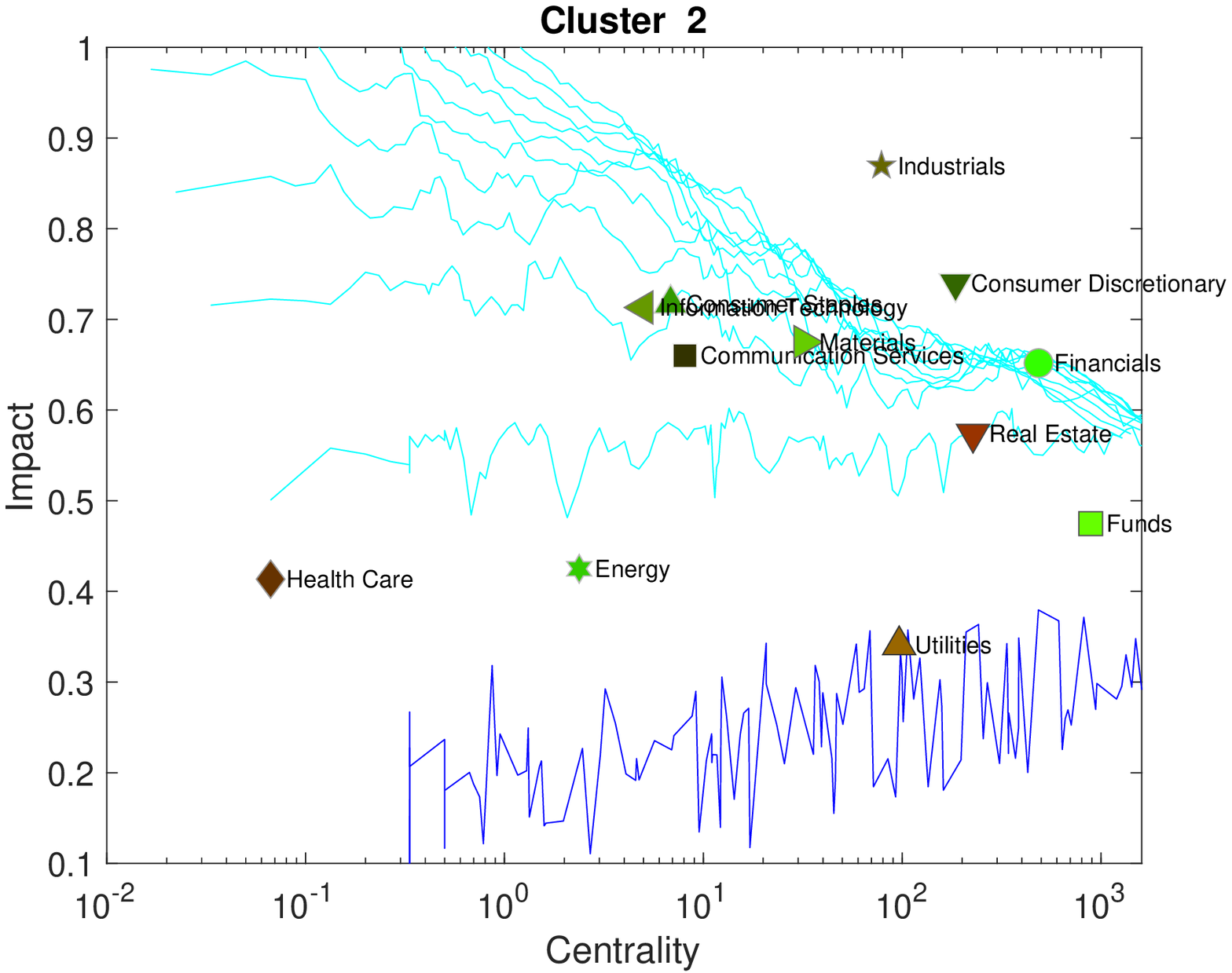}
\includegraphics[height=4.cm]{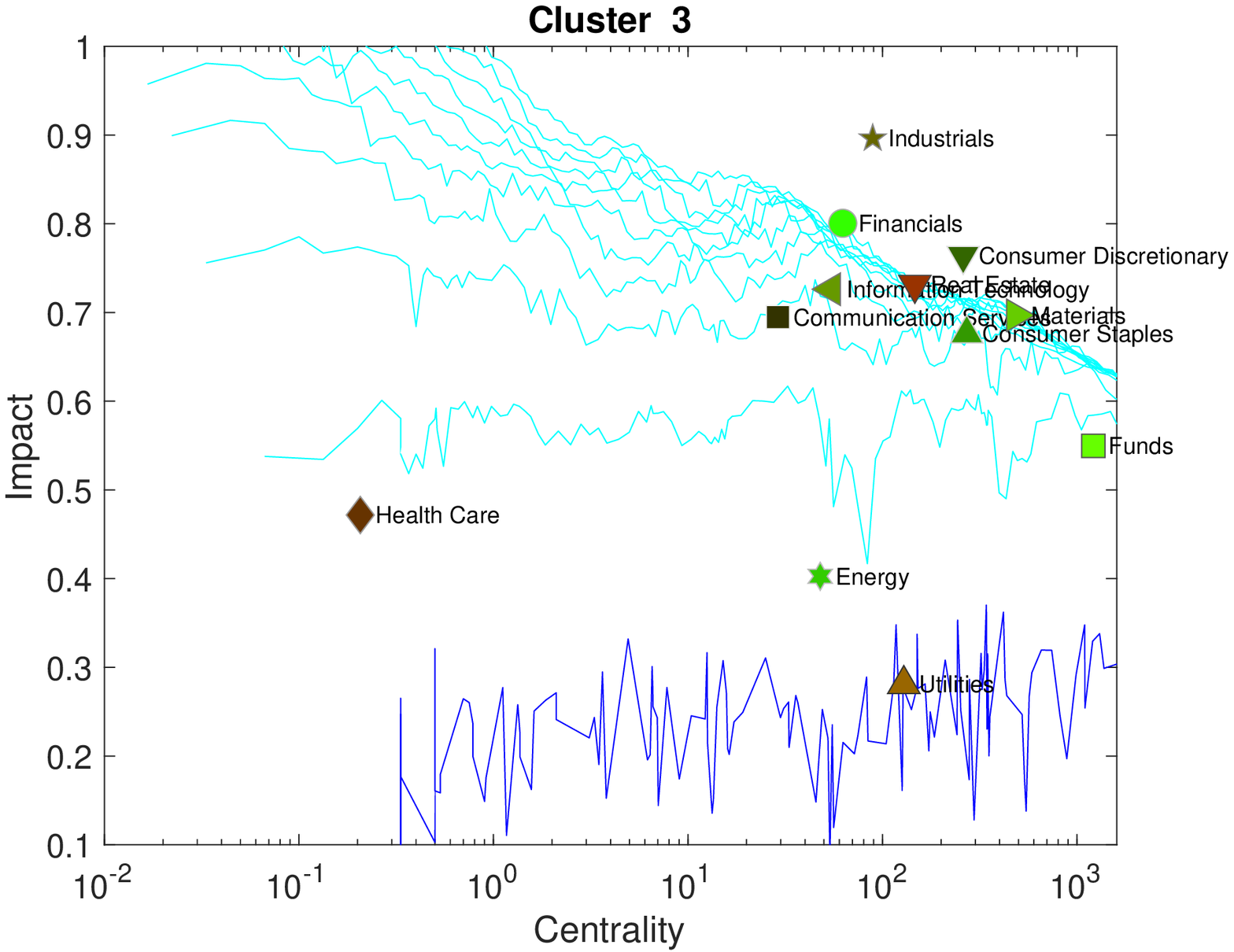}\\
\includegraphics[height=4.cm]{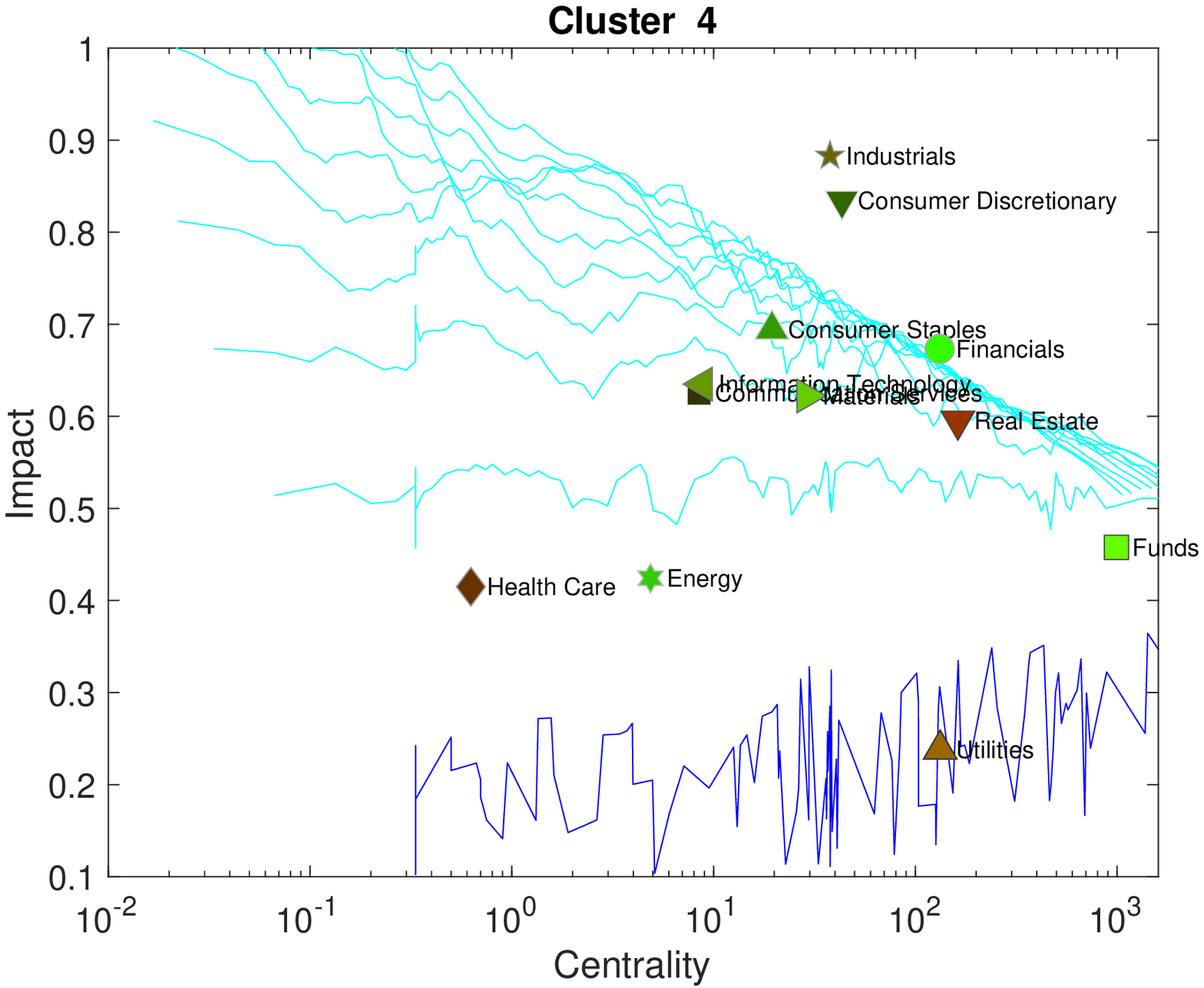}
\includegraphics[height=4.cm]{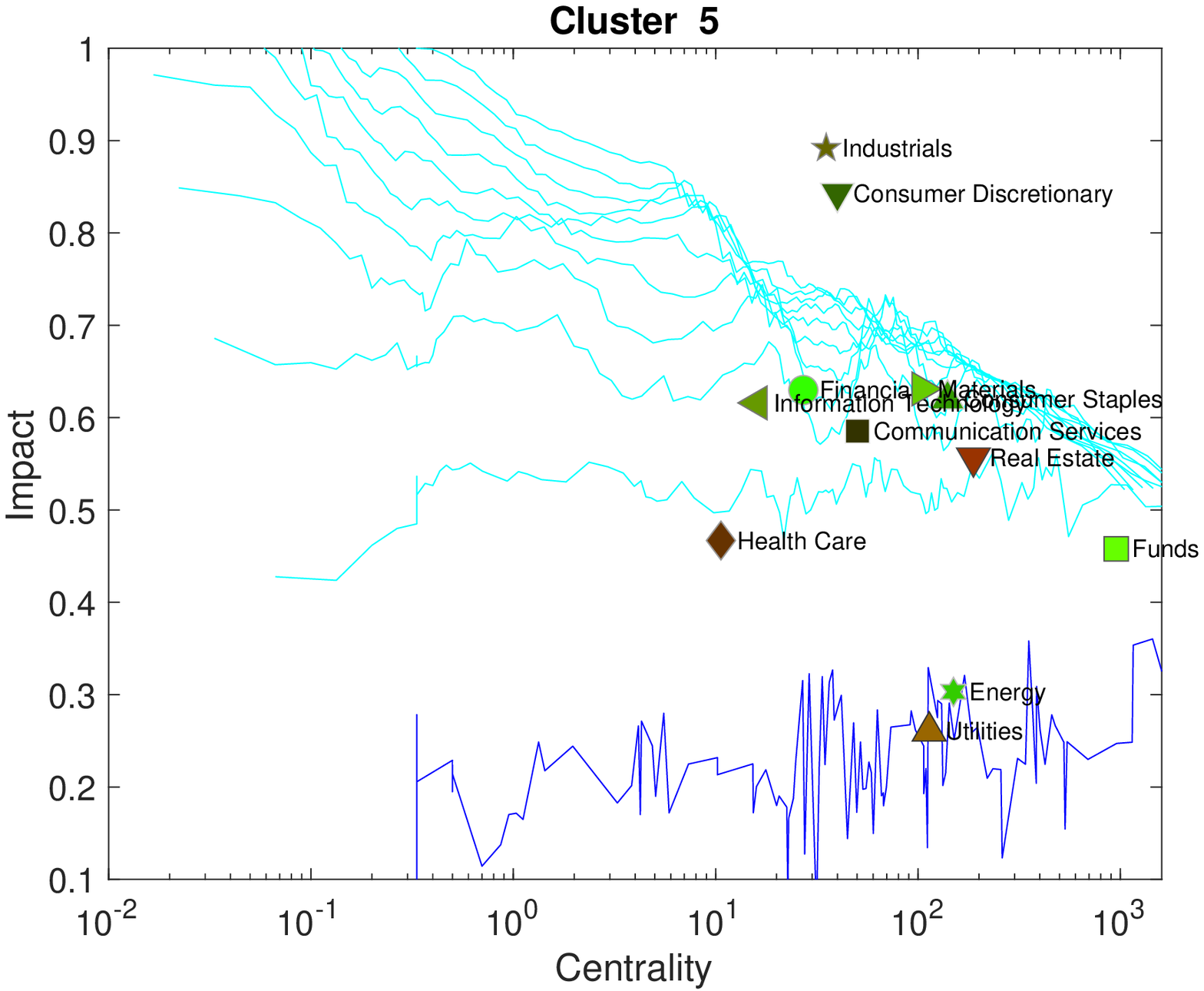}
\includegraphics[height=4.cm]{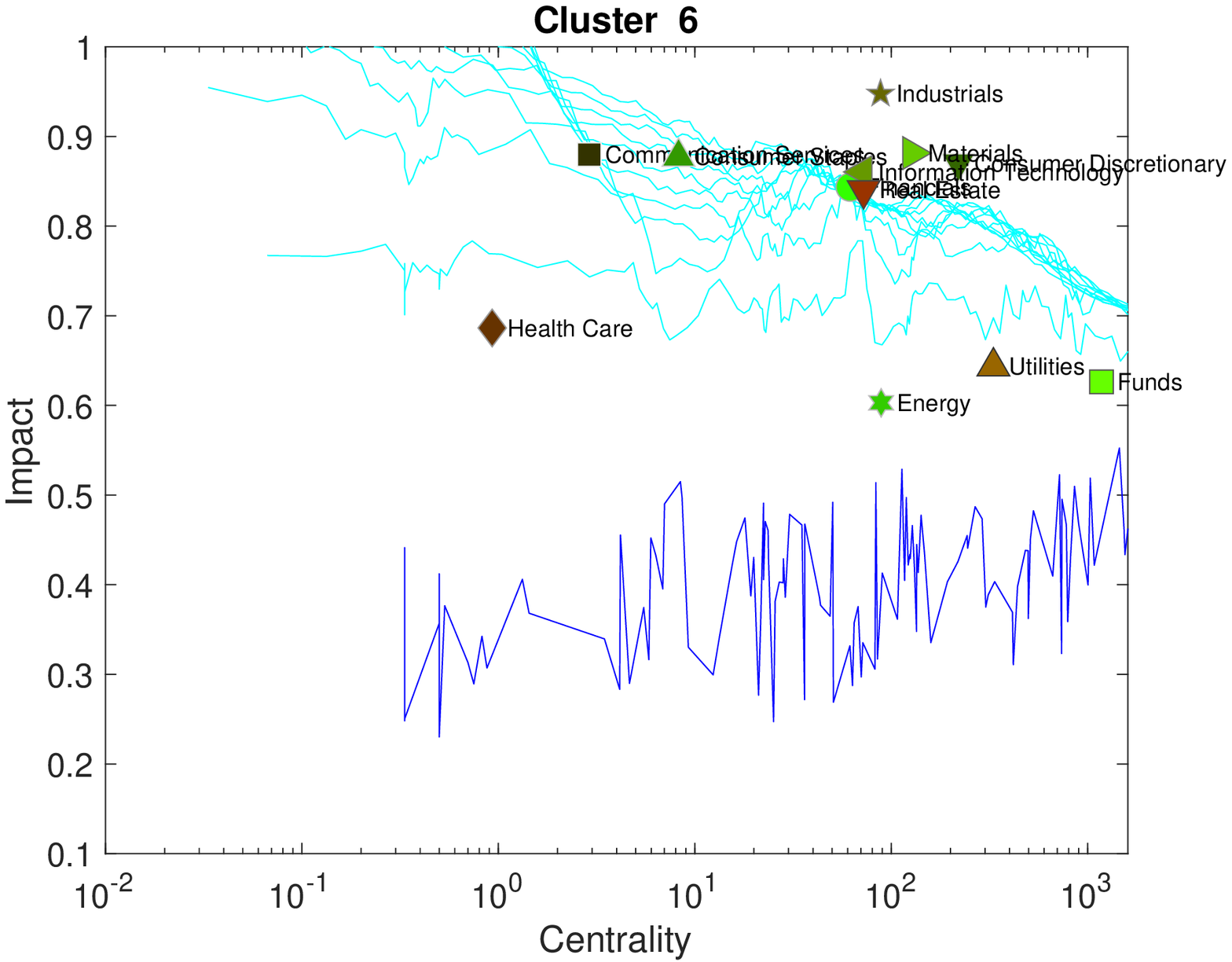}
\end{center}
\caption{Impact vs. centrality for the 6 clusters. 
The blue line corresponds to single nodes, cyan lines are group of nodes form 5 to 50.
Symbols are the supersectors.
}\label{ImpactVSCentraliatyMarketStates1}
\end{figure}

\begin{figure}[ht]
\begin{center}
\includegraphics[height=4.cm]{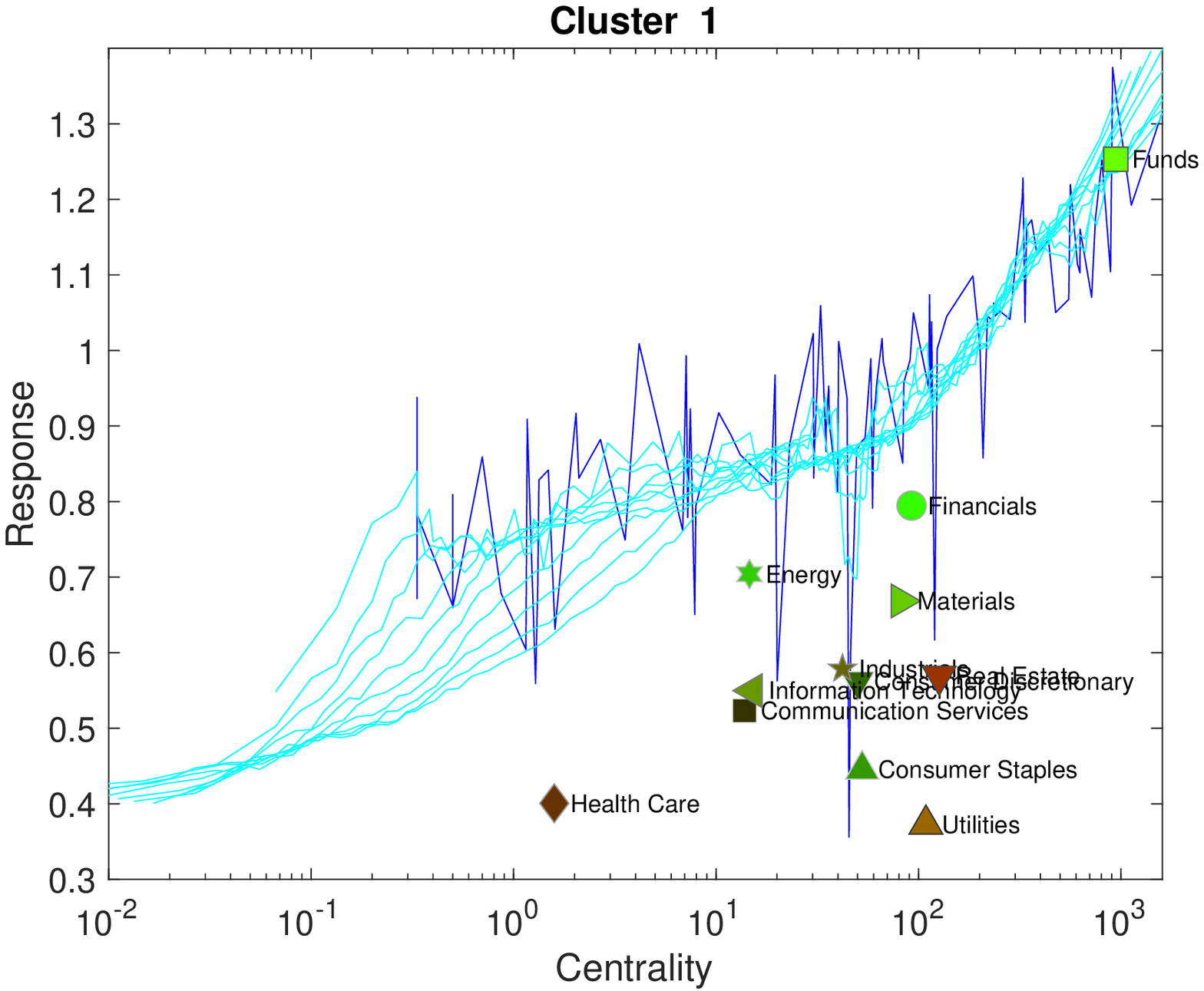}
\includegraphics[height=4.cm]{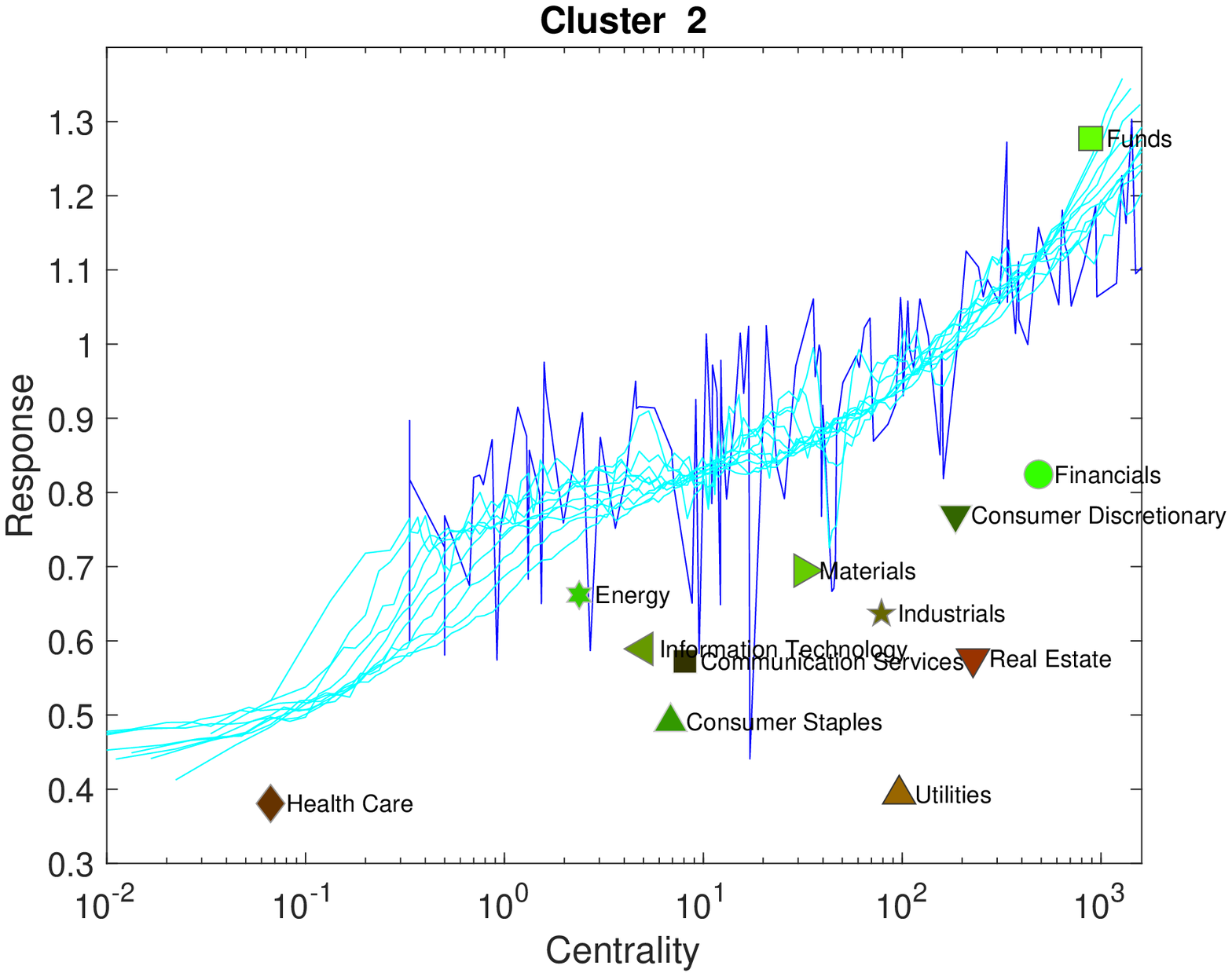}
\includegraphics[height=4.cm]{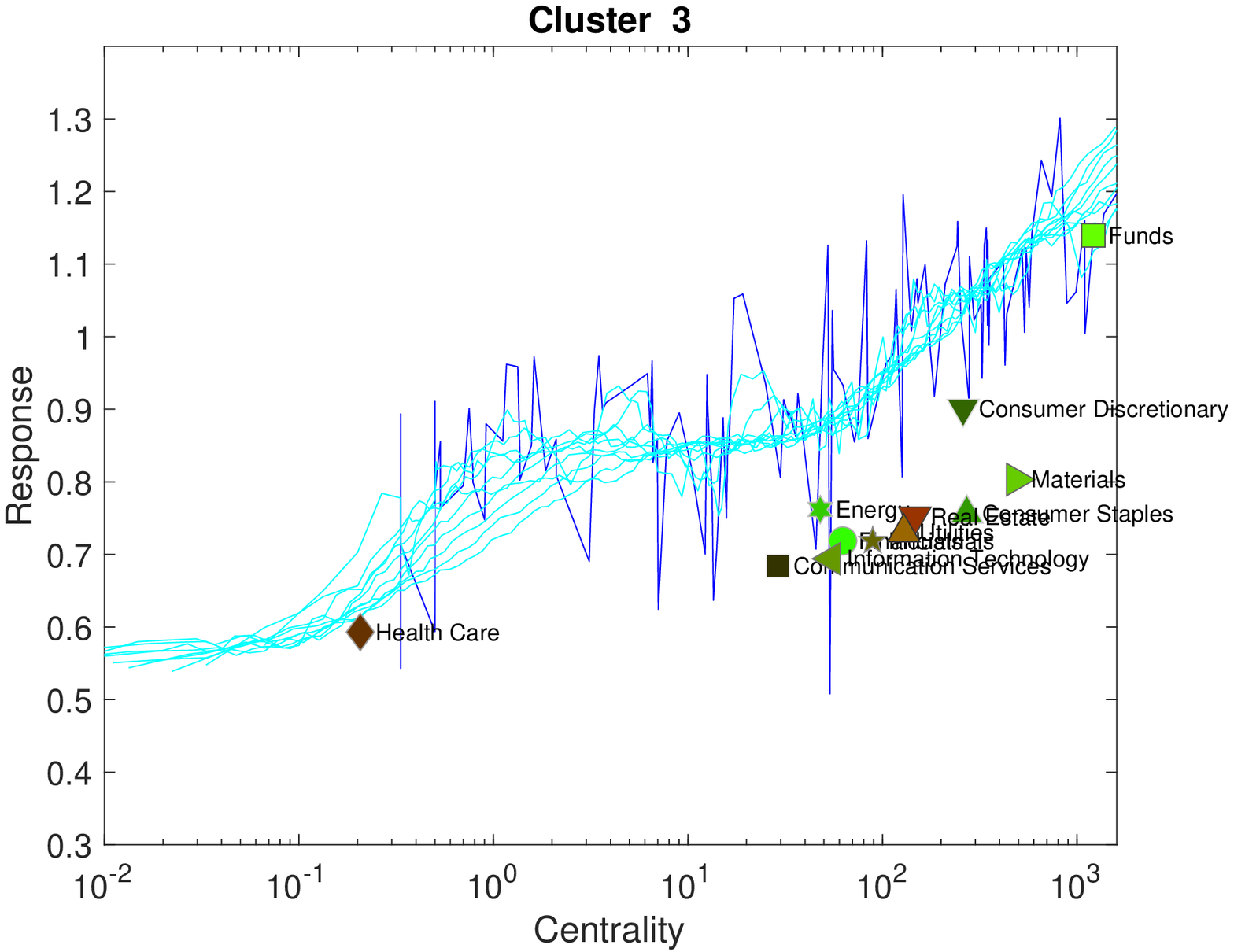}\\
\includegraphics[height=4.cm]{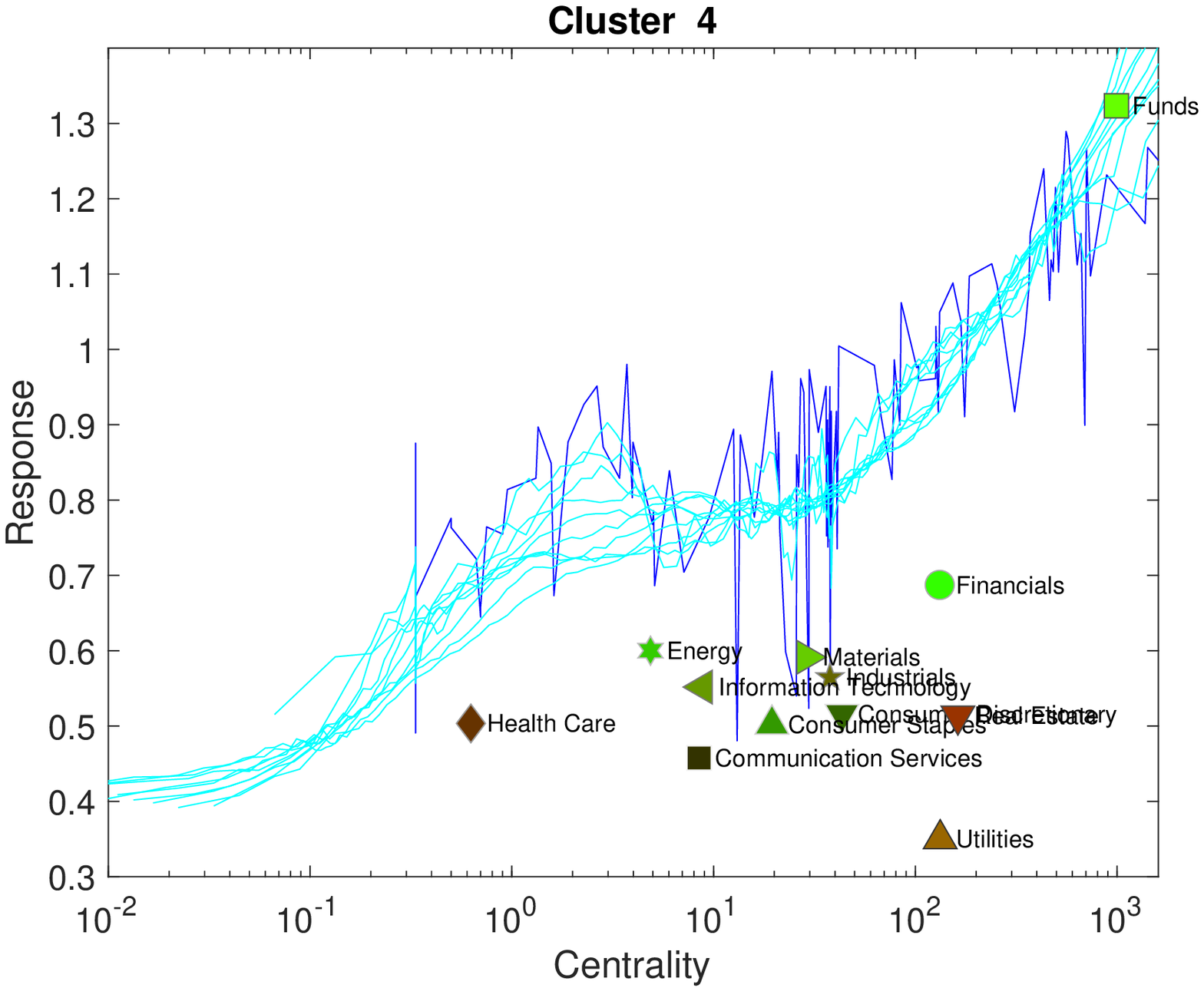}
\includegraphics[height=4.cm]{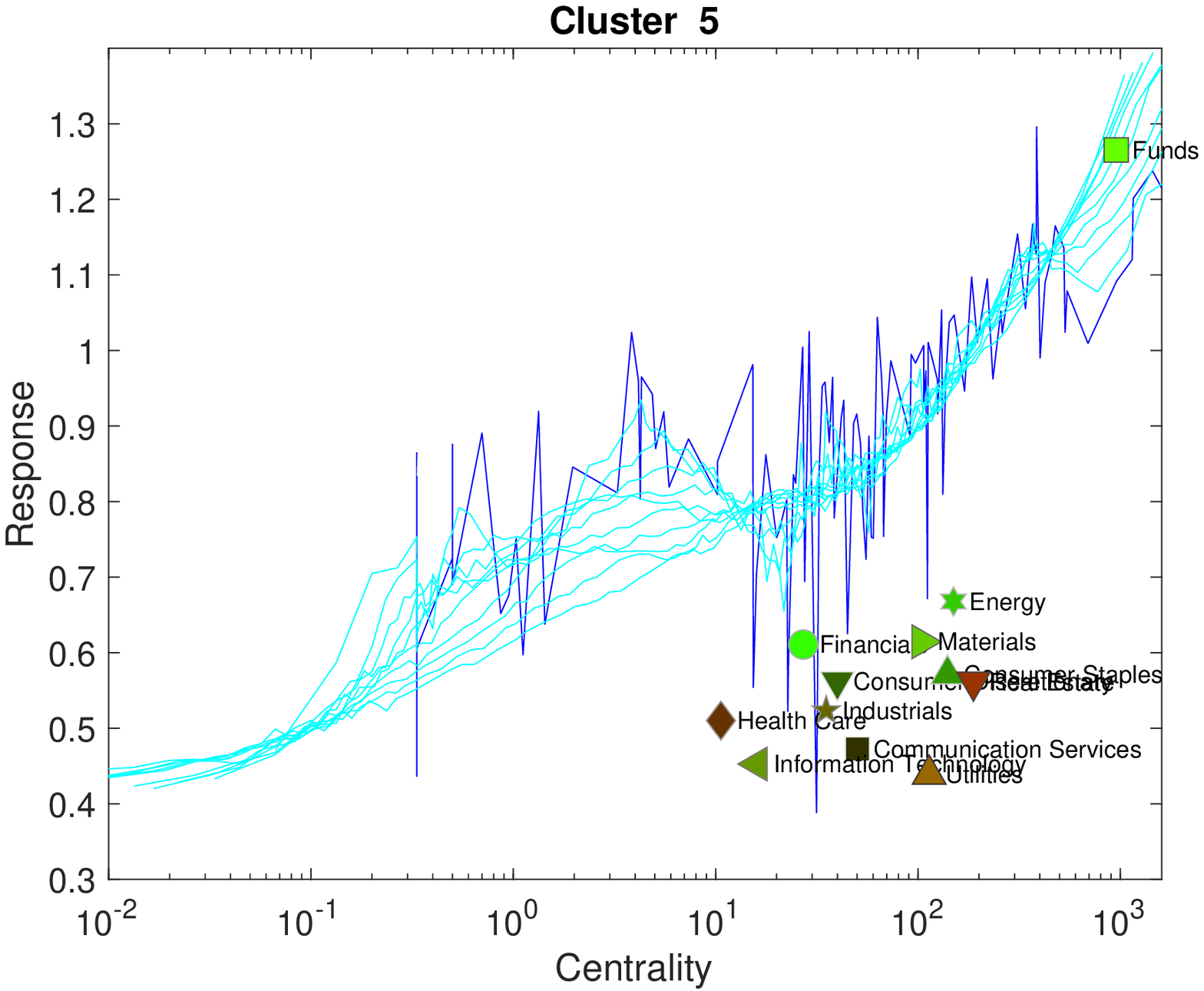}
\includegraphics[height=4.cm]{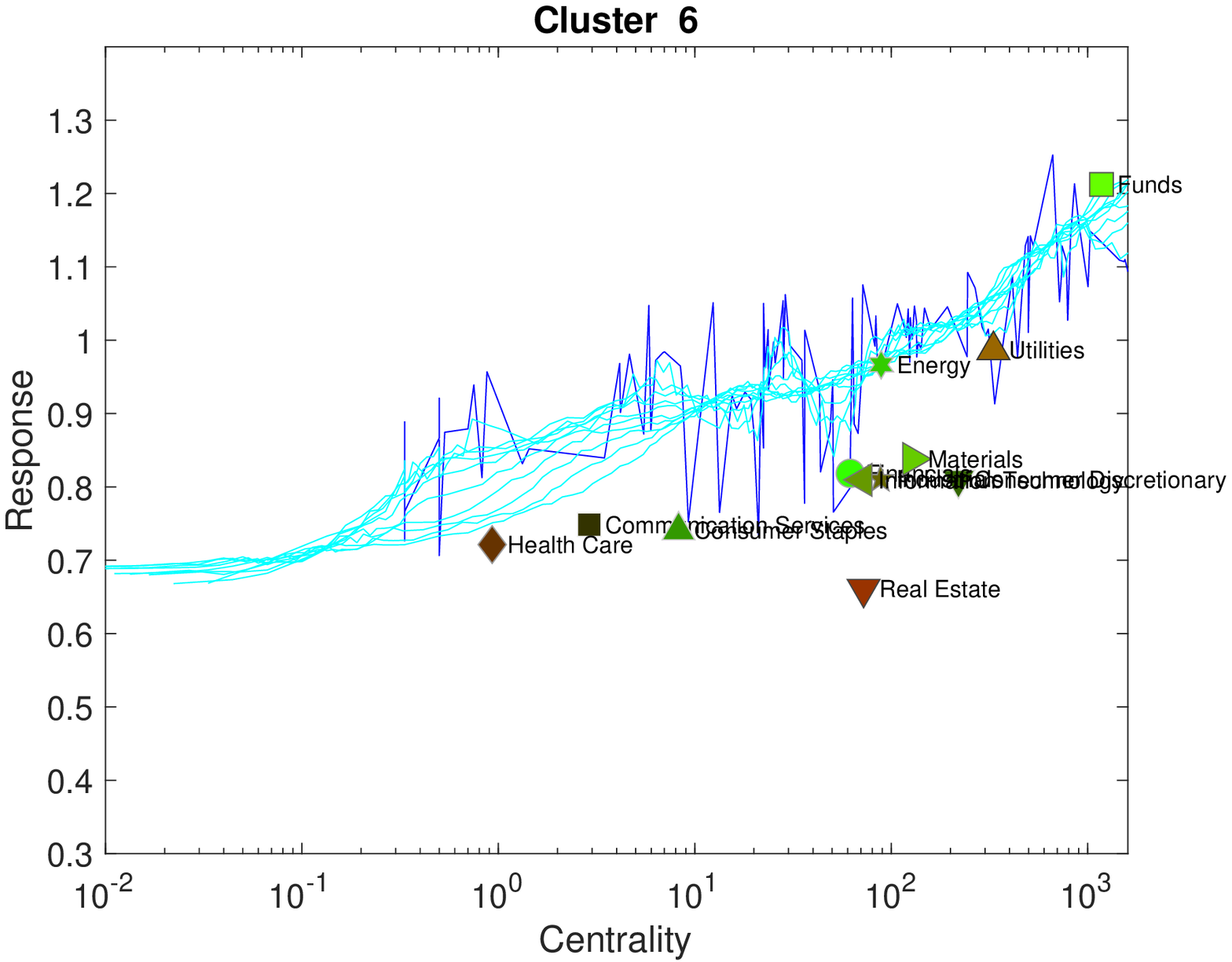}
\end{center}
\caption{Impacted vs. centrality for the 6 clusters. 
Blu line correspond to single nodes, cyan lines are group of nodes form 5 to 50.
Symbols are the supersectors.
}\label{ImpactVSCentraliatyMarketStates2}
\end{figure}

\subsection{Response and impact versus network centrality}\label{impactCentralityNodes}
We investigate the relation between impact and response of stocks with respect to their centrality within the information filtering network.
At the level of single nodes we observe that more central nodes have both higher impact and response. 
However, when we look at groups of nodes we observe that the most central clusters have a higher response but lower impact. Here the group centrality is simply the mean centrality of the group constituents.
This effect is most likely a consequence of the fact that central clusters are compact and therefore have a large internal effect on each other - which is not accounted for in the impact measure, and are less impactful on the rest of the system. 

This is illustrated in figure \ref{ImpactVSCentraliatyWP}\bc{,} where we report impact/response as function of the centrality of single nodes (blue lines) and the average  impact/response vs. average centrality for random groups of nodes with different sizes (cyan lines). These data refer to the whole period.
Specifically, in the top plot in  figure \ref{ImpactVSCentraliatyWP}, the lines show that centrality increases the response effect with little influence from the grouping.
Conversely, the bottom plot of  figure  \ref{ImpactVSCentraliatyWP} demonstrates that single nodes (blue lines) are also increasing their impact with centrality but when grouped the opposite happens and greater centrality corresponds to smaller impact (cyan lines).  
In these figures symbols report the average   impact/response vs. average centrality for the super-sectors. 
We see an overall consistency with the results for random groupings however with the sectors being overall a bit less responsive and impactful than the random groupings. 
 The analyses over the market states reported in Figures \ref{ImpactVSCentraliatyMarketStates1} and \ref{ImpactVSCentraliatyMarketStates2} give similar overall results with some significant differences in different market periods. 
When considering the impact scores, we see a clear increase in impact in cluster 6, both in the single and grouped node trends. We also see this to some extent in cluster 3, but this is limited to only some sectors. A more evident change is seen in the response scores of the supersectors in the two periods of crisis, which are dominated by market states three and six. 
In both periods of crisis, the funds sector has a lower response, with the majority of the other sectors seeing an upward shift in response score. In all cases, the sector specific response scores are seen to move closer to the single node and grouped node trend. Some sectors are more affected in times of crises than others, for example the Utilities and Healthcare sectors show larger increases in response and centrality. We also see a large increase in centrality on average across sectors in cluster 5.

\begin{figure}[ht]
\begin{center}
\includegraphics[width=0.8\textwidth]{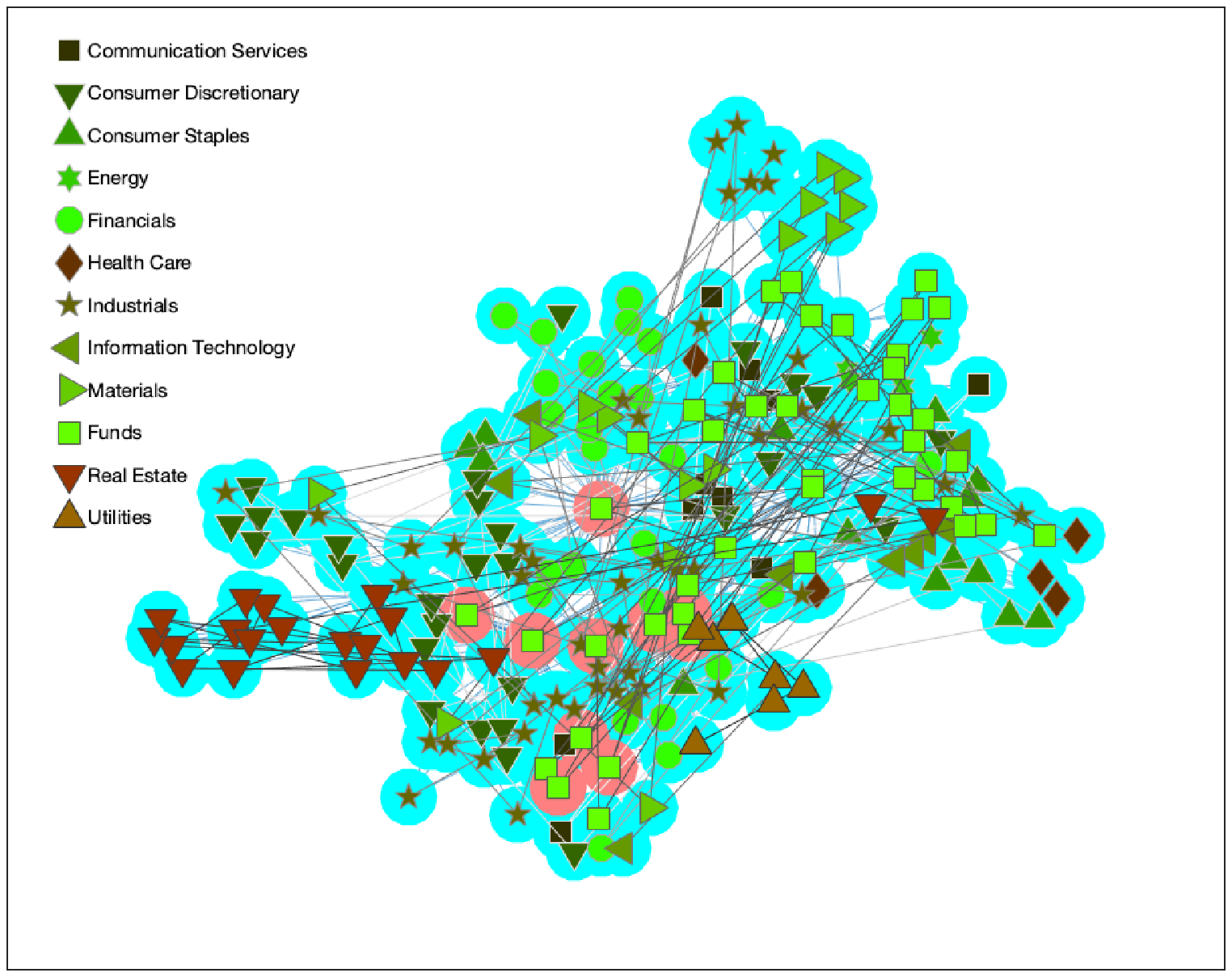} 
\end{center}
\caption{The group of 10 stocks with the collective highest impact computed over the full-period. 
}\label{MostImpactingWP}
\end{figure}

\subsection{Response and impact of supersectors}
The results of section \ref{impactCentralityNodes} show that, while more central nodes have both higher response and impact, if we aggregate nodes into random groups more central groups have higher response, but lower impact. Results are less clear when we group nodes into supersectors. 

Here we take a closer look to the aggregation of firms into supersectors, and we perform a regression analysis to understand if the impact and response measures of supersectors are affected by their centrality. 

There are two apparent differences between the random groups of sections \ref{impactCentralityNodes} and supersectors: First, random groups, associated with each of the cyan lines have all the same size, while the size of supersectors ranges from 4 to 47. Secondly, nodes in random groups tend to be dispersed in the network, while nodes within the same supersector tend to be closer to each other.
In order to account for this, we run a regression where, in addition to considering centrality as an independent variable, we control for the size of the supersector and  the fraction of a supersector's links that are shared by nodes within the supersector.  

The results are reported for one realization of the network in tables \ref{regressionTableImpacting} and \ref{regressionTableImpacted} for impact and response measures respectively.

\begin{table}[ht]
\begin{center}
\begin{tabular}{ l l l}
\hline
 \bf variable & \bf value & \bf p-value \\ 
\hline

 size & { }$1.33$ & { }$1.2\times 10^{-12}$ \\  
 fraction of links within supersector & $-1.06$ & { }$7\times 10^{-8}$\\  
 log centrality & $-0.02$ & { }$0.86$\\    
\hline
\end{tabular}
\end{center}
\caption{Linear regression for the measure of impact ($R^2=0.53$).}\label{regressionTableImpacting}
\end{table}

\begin{table}[ht]
\begin{center}
\begin{tabular}{ l r r}
\hline
 \bf variable & \bf value & \bf p-value \\ 
\hline
 size & $-0.34$ & $0.03$ \\  
 fraction of links within supersector & $0.79$ & $3\times 10^{-5}$\\  
 log centrality & $0.25$ & $0.026$\\    
\hline
\end{tabular}
\end{center}
\caption{Linear regression for the response measure ($R^2=0.52$).}
\label{regressionTableImpacted}
\end{table}

We see from table \ref{regressionTableImpacting} that the coefficients associated with size and fraction of links shared by nodes of a supersector are statistically significant. The positive sign of the coefficient corresponding to the size is simply due to the fact that as the size of the supersector increases the shock affects a higher number of nodes. The negative sign of the fraction of links connecting nodes within the supersector is instead due to the fact that links between nodes that are exogenously shocked do not increase the impact, so that the more compact a supersector is, the less it affects the others. We also observe that the centrality of supersectors does not appear to play a significant role here, and that the effect of the network structure is simply accounted for by the fraction of links within supersectors.

This is not the case for the response measure. We see from table \ref{regressionTableImpacted} that centrality remains statistically significant (p-value 2.6\%, see table \ref{regressionTableImpacted}) with a positive coefficient, signaling the tendency of more central sectors to show a higher response. We also observe a change in the sign of the coefficients associated with the other two variables. In particular we see that the fraction of links shared by nodes of a supersector now has a positive coefficient, denoting the fact that more compact supersectors tend to have a higher response. This is due to the fact that, at odds with the case of impact, the contribution of these links to loss propagation is accounted for in the calculation of the response variable.

\subsection{Nodes with the highest impact}
Concerning the identification of the nodes with the highest impact, we find that the most influential  $n=10$ nodes  for the whole period are:
MRC	, 	
WTA	, 	
CTY,	 	
EDI,		ASL,		 BRS,		HSL,		 
TMP,		 
PLI,		 FSV, which all belong to the funds supersector.
Their position in the network is highlighted in figure \ref{MostImpactingWP}.  
We see that these stocks don't localise to a particular region of the network, suggesting that they are not necessarily displaying similar behaviours leading to this higher level of impact.
Note that this is the group of 10 stocks with collective highest impact, not the collection of single highest impacting stocks.

We observe similar results across the different market states, as summarized in table \ref{tab:my_label}. Firstly, we note again that the majority these stocks belong to the funds supersector, suggesting that stocks with a high level of diversification are likely to impact the rest of the market the most when experiencing shocks. 
We also notice that there are several repetitions in the groups of highest impact stocks across the six different periods.
For instance, CTY appears in all six clusters and there are five other stocks (MRC, EDI, ASL, TMP, PLI) that appear in five out of six clusters. They also all appear in the full-period group reported in figure \ref{MostImpactingWP}.
Cluster 3 is the one with least overlap with the rest of the clusters indicating that the risk scenario during the 2007-2009 crisis was quite special.
Conversely, Cluster 6 has similar overlaps as the other clusters suggesting that the 2020 covid-related crisis is instead less peculiar.

 \begin{table}
     \centering
     \begin{tabular}{|c|c|c|c|c|c|}
     \hline
       Cluster 1   & Cluster 2 & Cluster 3 & Cluster 4 & Cluster 5 & Cluster 6 \\
       \hline
        ATS  & CI & FCI & FGT & SDR** & ICP** \\
        MRC & ATS & ATS & MRC & FGT & MRC \\
        WTA & TEM & MRC & CTY & MRC & CTY \\
        CTY & CTY & WTA & TRY & WTA & TRY \\
        TRY & MYI & CTY & EDI & CTY & EDI \\
        EDI & JMG & FEV & ASL & TRY & ASL \\
        ASL & FEV & CLD & HSL & EDI & BRS \\
        AGT & EDI & ASL & TMP & ASL & HSL \\
        TMP & TMP & IMI* & PLI & TMP & TMP \\
        PLI & PLI & LWD** & FSV & PLI & PLI \\
         \hline
     \end{tabular}
     \caption{
Groups of 10 stocks with highest impact across the six different time-clusters.     
     *Capital Goods	Industrials, 
**Diversified Financials, all others Funds}
     \label{tab:my_label}
 \end{table}

\section{Conclusion}
In this paper, we have presented a novel method to measure the impact and response of stocks in a market when shocks are experienced.  For the first time, we associate the structure of the information filtering network with a quantitative risk measure relating the behavior of impact and response to the structure of the underlying inverse covariance matrix of the stock log-returns. In the application to different market states observed from 2005 to 2020 in FTSE 100 and 250 markets, we observe that both impact and response are related to the centrality of individual nodes in the network, but that central groups have a higher response but lower impact due to the internal effects for these groups.  

We observe markedly different behavior in different states of the market, particularly in the 2008 and COVID-19 crises. The convergence to the single and grouped node trend observed for response scores during periods of crisis is consistent with observations that correlations between all stock prices are seen to increase in these periods. This is not only consistent with our own observations but also with recent observations made by Sandoval et. al. \cite{Leonidas_2011}. This tendency of `markets to behave as one' in times of crisis is interesting to monitor from the context of response and impact scores - particularly in the case of sectors which, in a period of comparative market stability, are seen to have a lower value according to response than the individual/group trend. These sectors might be expected to respond worse to the crisis period, whereas the effect on sectors closer to the trend, or above, may see a lesser impact. 

Finally, we observe that the most impactful stocks in the markets are belonging to funds. This suggests that stocks with a high level of diversification are more likely to present a larger knock-on effect to the rest of the market when they experience shocks. 

\section*{Acknowledgments}\noindent
The author acknowledges discussions with many members of the Financial Computing and Analytics group at UCL. In particular a special thank to Guido Massara, Carolyn Phelan and Pier Francesco Procacci.  
Also, thanks for support from ESRC (ES/K002309/1),  EPSRC (EP/P031730/1) and EC (H2020-ICT-2018-2 825215).

\bibliographystyle{unsrt}
\bibliography{StressTestMultivar.bib}

\begin{thebibliography}{10}

\bibitem{aste2020stress}
Tomaso Aste.
\newblock Stress testing and systemic risk measures using multivariate
  conditional probability.
\newblock {\em Available at SSRN 3575512 and https://arxiv.org/pdf/2004.06420},
  2020.

\bibitem{pozzi2013spread}
Francesco Pozzi, Tiziana Di~Matteo, and Tomaso Aste.
\newblock Spread of risk across financial markets: better to invest in the
  peripheries.
\newblock {\em Scientific reports}, 3:1665, 2013.

\bibitem{procacci2019forecasting}
Pier~Francesco Procacci and Tomaso Aste.
\newblock Forecasting market states.
\newblock {\em Quantitative Finance}, 19(9):1491--1498, 2019.

\bibitem{Leonidas_2011}
Leonidas Sandoval and Italo Franca.
\newblock Correlation of financial markets in times of crisis.
\newblock {\em Physica A: Statistical Mechanics and its Applications}, 391, 02
  2011.

\bibitem{BOEFSR2020}
A.~Bailey, J.~Cunliffe, B.~Broadbent, D.~Ramsden, S.~Woods, C.~Woolard,
  A.~Brazier, B.~Colette, A.~Kashyap, D.~Kohn, Stheemanm E., and C.~Roxburgh.
\newblock Financial stability report.
\newblock {\em Bank of England Financial Policy Committee}, page~45, 2020.

\bibitem{IMF_stress}
R.~Anderson, J.~Danielsson, C.~Baba, U.S. Das, H.~Kang, and Segoviano M.
\newblock Macroprudential stress tests and policies: Searching for robust and
  implementable frameworks.
\newblock {\em IMF Working Paper}, page WP/18/197, 2018.

\bibitem{tumminello2005tool}
Michele Tumminello, Tomaso Aste, Tiziana Di~Matteo, and Rosario~N Mantegna.
\newblock A tool for filtering information in complex systems.
\newblock {\em Proceedings of the National Academy of Sciences},
  102(30):10421--10426, 2005.

\bibitem{aste2010correlation}
Tomaso Aste, W~Shaw, and Tiziana Di~Matteo.
\newblock Correlation structure and dynamics in volatile markets.
\newblock {\em New Journal of Physics}, 12(8):085009, 2010.

\bibitem{markose2012too}
Sheri Markose, Simone Giansante, and Ali~Rais Shaghaghi.
\newblock ‘too interconnected to fail’financial network of us cds market:
  Topological fragility and systemic risk.
\newblock {\em Journal of Economic Behavior \& Organization}, 83(3):627--646,
  2012.

\bibitem{battiston2012debtrank}
Stefano Battiston, Michelangelo Puliga, Rahul Kaushik, Paolo Tasca, and Guido
  Caldarelli.
\newblock Debtrank: Too central to fail? financial networks, the fed and
  systemic risk.
\newblock {\em Scientific reports}, 2(1):1--6, 2012.

\bibitem{martinez2014empirical}
Serafin Martinez-Jaramillo, Biliana Alexandrova-Kabadjova, Bernardo
  Bravo-Benitez, and Juan~Pablo Sol{\'o}rzano-Margain.
\newblock An empirical study of the mexican banking system’s network and its
  implications for systemic risk.
\newblock {\em Journal of Economic Dynamics and Control}, 40:242--265, 2014.

\bibitem{bravo2016centrality}
Bernardo Bravo-Benitez, Biliana Alexandrova-Kabadjova, and Serafin
  Martinez-Jaramillo.
\newblock Centrality measurement of the mexican large value payments system
  from the perspective of multiplex networks.
\newblock {\em Computational Economics}, 47(1):19--47, 2016.

\bibitem{bruyckere2015ecb}
V.~De~Bruyckere.
\newblock Systemic risk rankings and network cnetrality in the european banking
  sector.
\newblock {\em ECB working paper}, 2015.

\bibitem{kuzubas2014turkishfc}
T.~U. Kuzubaş and B.~Ömercikoğlu, I.and~Saltoğlu.
\newblock Network centrality measures and systemic risk: An application to the
  turkish financial crisis.
\newblock {\em Physica A}, pages 203--215, 2014.

\bibitem{billio2012econometric}
Monica Billio, Mila Getmansky, Andrew~W Lo, and Loriana Pelizzon.
\newblock Econometric measures of connectedness and systemic risk in the
  finance and insurance sectors.
\newblock {\em Journal of financial economics}, 104(3):535--559, 2012.

\bibitem{diebold2014topology}
F.~X. Diebold and K.~Yilmaz.
\newblock On the network topology of variance decompositions: Measuring the
  connectedness of financial firms.
\newblock {\em Journal of Econometrics}, pages 119--134, 2014.

\bibitem{LoGo16}
Wolfram Barfuss, Guido~Previde Massara, Tiziana Di~Matteo, and Tomaso Aste.
\newblock Parsimonious modeling with information filtering networks.
\newblock {\em Physical Review E}, 94(6):062306, 2016.

\bibitem{aste2020topological}
Tomaso Aste.
\newblock Topological regularization with information filtering networks.
\newblock {\em arXiv preprint arXiv:2005.04692}, 2020.

\bibitem{massara2019learning}
Guido~Previde Massara and Tomaso Aste.
\newblock Learning clique forests.
\newblock {\em arXiv preprint arXiv:1905.02266}, 2019.

\bibitem{massara2016network}
Guido~Previde Massara, Tiziana Di~Matteo, and Tomaso Aste.
\newblock Network filtering for big data: Triangulated maximally filtered
  graph.
\newblock {\em Journal of complex Networks}, 5(2):161--178, 2016.

\bibitem{nicola2020information}
Giancarlo Nicola, Paola Cerchiello, and Tomaso Aste.
\newblock Information network modeling for us banking systemic risk.
\newblock {\em Entropy}, 22(11):1331, 2020.

\bibitem{turiel2020simplicial}
Jeremy~D Turiel, Paolo Barucca, and Tomaso Aste.
\newblock Simplicial persistence of financial markets: filtering, generative
  processes and portfolio risk.
\newblock {\em arXiv preprint arXiv:2009.08794}, 2020.

\bibitem{christensen2018network}
Alexander~P Christensen, Yoed~N Kenett, Tomaso Aste, Paul~J Silvia, and
  Thomas~R Kwapil.
\newblock Network structure of the wisconsin schizotypy scales--short forms:
  Examining psychometric network filtering approaches.
\newblock {\em Behavior Research Methods}, 50(6):2531--2550, 2018.

\bibitem{friedman2008sparse}
Jerome Friedman, Trevor Hastie, and Robert Tibshirani.
\newblock Sparse inverse covariance estimation with the graphical lasso.
\newblock {\em Biostatistics}, 9(3):432--441, 2008.

\bibitem{tong1978TAR}
H.~Tong.
\newblock On a threshold model.
\newblock {\em Pattern Recognition and Signal Processing}, 1978.

\bibitem{ren2017pattern}
L.~Ren, Y.~Wei, J.~Cui, and Y.~Du.
\newblock A sliding window-based multi-stage clustering and probabilistic
  forecasting approach for large multivariate time series data.
\newblock {\em Journal of Statistical Computation and Simulation}, pages
  2494--2508, 2017.

\bibitem{nevill1997hierarchical}
C.G. Nevill-Manning and I.H. Witten.
\newblock Identifying hierarchical structure in sequences: A linear-time
  algorithm.
\newblock {\em Journal of Artificial Intelligence Research}, pages 67--82,
  1997.

\bibitem{liao2005dtw}
W.T. Liao.
\newblock Clustering of time series data-a survey.
\newblock {\em Pattern Recognition}, pages 1857--1874, 2005.

\bibitem{procacci2020covidstates}
P.~Procacci, C.~Phelan, and T.~Aste.
\newblock Market structure dynamics during covid-19 outbreak.
\newblock {\em arXiv:2003.10922v1 [q-fin.ST]}, 2020.

\bibitem{pharasi2020states}
H.~K. Pharasi, E.~Seligman, and T.~H. Seligman.
\newblock Market states: A new understanding.
\newblock {\em http://arxiv.org/pdf/2003.07058}, Revised 2020.

\bibitem{munnix2012states}
M.~Münnix, T.~Shimada, R.~Schäfer, F.~Leyvraz, T.H. Seligman, T.~Guhr, and
  Stanley H.
\newblock Identifying states of a financial market.
\newblock {\em Scientific Reports}, 2(644), 2012.

\bibitem{hendricks2020potts}
D.~Hendricks, T.~Gebbie, and D.~Wilcox.
\newblock Detecting intraday financial market states using temporal clustering.
\newblock {\em Quantative Finance}, 16(11):1657--1678, 2015.

\bibitem{TMFG}
Guido~Previde Massara, T.~Di~Matteo, and Tomaso Aste.
\newblock Network filtering for big data: Triangulated maximally filtered
  graph.
\newblock {\em Journal of Complex Networks}, 5(2):161, 2017.

\bibitem{procacci2021portfolio}
Pier~Francesco Procacci and Tomaso Aste.
\newblock Portfolio optimization with sparse multivariate modelling.
\newblock {\em arXiv preprint arXiv:2103.15232}, 2021.

\end{thebibliography}

\end{document}